# Data-mining for Fault-Zone Detection of Distance Relay in FACTS-Based Transmission

Nima Salek Gilani
*Department of Electrical Engineering*
K.N.Toosi University of Technology
Tehran, Iran
Nima.salek@email.kntu.ac.ir

Mohammad Tavakoli Bina
*Department of Electrical Engineering*
K.N.Toosi University of Technology
Tehran, Iran
Tavakoli@eetd.kntu.ac.ir

Fatemeh Rahmani
*Department of Electrical Engineering*
Lamar University
Beaumont, TX, USA
frahmani@lamar.edu

Mahmood Hosseini Imani
*Department of Energy DENERG*
Politecnico di Torino
Turin, Italy
mahmood.hosseiniimani@polito.it

*Abstract*—In this study, the problem of fault zone detection of distance relaying in FACTS-based transmission lines is analyzed. Existence of FACTS devices on the transmission line, when they are included in the fault zone, from the distance relay point of view, causes different problems in determining the exact location of the fault by changing the impedance seen by the relay. The extent of these changes depends on the parameters that have been set in FACTS devices. To solve the problem associated with these compensators, two instruments for separation and analysis of three line currents, from the relay point of view at fault instance, have been utilized. The wavelet transform was used to separate the high-frequency components of the three line currents, and the support vector machine (using methods for multi-class usage) was used for classification of fault location into three protection regions of distance relay. Besides, to investigate the effects of TCSC location on fault zone detection of distance relay, two places, one in fifty percent of line length and the other in seventy-five percent of line length, have been considered as two scenarios for confirmation of the proposed method. Simulations indicate that this method is effective in the protection of FACTS-based transmission lines.

*Keywords— Data-mining, Distance relay, FACTS, SVM*

## I. Introduction

Increasing demand, caused by the development of cities and growing population, draws more attention on power transmission limits. Power Electronics helps to solve the problem by introducing flexible ac transmission system (FACTS) devices. The general advantages of FACTS devices are well known. However, FACTS devices can lead to severe problems in determining the exact location of the fault when they are included in the fault zone. Distance relay's function is based on measured impedance by which they calculate the fault location. Installed FACTS devices on the transmission line change the impedance measured by distance relay, which consequently leads to wrong fault zone detection. The amount of deviation in the impedance depends on the parameters settings of the FACTS device.

In recent years, researchers have become increasingly interested in solving this issue. For modeling series capacitor in the fault period, most of the papers used the model presented by Goldsworthy [1]. In [2], a solution based on time equations of network with consideration of an extensive model of a transmission line is presented. In this solution, zone and location of the fault, independent of the fault resistance, have been estimated using two subroutines. Some papers developed their methods using measurements from principle components of the main frequency. Saha et al. in [3] developed a method based on components of main frequency on one side of the transmission line using the Goldsworthy model.

In [4], an algorithm based on forward and backward waves is presented to protect series compensated transmission lines; although only one phase to ground fault has been analyzed in this paper.

Data have been presented in the literature [5-7] which suggest methods based on neural networks. It has been suggested to use RBFN as a radial neural network to classify fault type and fault location [7]. Reference [6] trains two different types of neural networks to identify fault type and location. Also, [5] uses the EDBD algorithm to calculate the fault location. Neural networks have limitations, including their need for a massive amount of neurons to model network structure. Therefore, numerous data from the electrical network is needed, and also much time to train the network. Nevertheless, these papers used a limited number of data to test their methods.

Girgis et al. in [8] proposed a method based on Kalman filter [9] which estimate the state of a dynamic system from a series of measurements containing errors. In [10], a method using wavelet transform is suggested. In both [8] and [10], limited cases have been used to assess the effectiveness of the proposed methods. Furthermore, in [10] high sampling frequency of 200 KHz is used, which seems to be difficult in practical implementation.

In [11], a data mining method is used for three purposes. Support vector machine (SVM) has become a popular tool in the power grid [12]. In this paper, three separate SVM is trained to determine fault location, presence or absence of fault and whether the fault includes the ground or not. This reference uses samples of firing angle and current, starting from the fault instance till half cycle after it. Considering 400 kV and 50 Hz case study and sampling frequency of 1 KHz in this paper, ten samples of current from each phase is analyzed. In this paper, fault location is confined to before and after the compensator, and its effectiveness is studied only in 200 cases.

In [13], a two stage method is proposed to identify the fault location. In this paper, samples of three line currents for one cycle after the fault instance is used to train SVM. Considering 400 kV and 50 Hz case study and sampling frequency of 1 KHz in this paper, 80 samples of current from each phase is analyzed. In the first stage, high frequencies of three line currents are separated using wavelet transform; then in the second stage, these separated signals are given to SVM to classify the fault location to before and after the compensator. The problem is that the model used for simulation of the transmission line is not proper to study non-base frequencies while this paper used high frequencies to separate the fault locations. This method, considering using SVM, can only separate the fault location to before and after the compensator. Taking into account 400 kV and 50 Hz case study and sampling frequency of 1 KHz in this paper, 60 samples are analyzed.

Researchers in [14] proposed the random forest method [15-18] to identify the fault location. This data mining method is founded based on the decision tree (DT) method. In this method, several independent unpruned decision trees are produced using different subsets of the data. Each tree has one vote, and the most vote determines the result. Samples of current and voltage in half cycle after the fault instance are utilized for analysis. Seventy percent of the samples are used for the training process of random forest, and the remaining thirty percent is used for the testing process. Considering a large amount of training data, good results have been achieved. Similarly, in this paper, the fault location is limited to before and after the compensator.

The object of this research is to improve distance protection performance, using data mining methods while considering FACTS-based transmission lines. Earlier works using artificial neural network suffer from limitations on high number of neurons. These limitations can be overcome by using more efficient tools like support vector machines. Furthermore, a more precise simulation of the transmission line is conducted here using PSCAD/EMTDC software, as well as separating the fault areas into three zones according to distance relay settings. Besides, to investigate the effects of TCSC location on fault zone detection of distance relay, two places --one in fifty percent of line length and the other in seventy five percent of line length-- have been considered as two scenarios for confirmation of the proposed method. The simulations show a 94.2% of success rate in scenario one and 86.7% in the other, which are high success rates, which indicate that placing TCSC in fifty percent of line length leads to a higher prediction accuracy. Moreover, these success rates are achieved using only 15.2% of total data as the training set, which shows the high capability of proposed method.

## II. PROPOSED METHOD

In this study, PSCAD software is used for simulation. Three line currents are recorded after creating a variety of fault conditions on the simulated transmission line. These outputs from PSCAD are used as inputs to wavelet transform in order to extract high frequencies (between 1000 to 2000 Hz). It is reported in [19] and [20] that the most proper wavelet in studying faults in power system is Daubechies (db) wavelets. In [13], it was reported that from different types of db mother wavelets, db2 gives the best results in the detection of fault location, in the presence of series capacitor in a transmission line. It was also reported that frequencies between 1000 to 2000 Hz are suitable for best fault zone classification. Therefore, considering a sampling frequency of 4 KHz in PSCAD in this study, single-level discrete wavelet transform with db2 mother wavelet is carried out to obtain high frequency samples. After that, wavelet transform's outputs are used as inputs to SVM to create a hyper plane to separate fault zones.

## III. CASE STUDIES

### A. Simulation of case studies in PSCAD

For TCSC simulation in PSCAD, reference model in [13] is used. This model is shown in Fig. 1. The capacitance of SC capacitor changes in proportion with the firing angle.

The feasibility of the proposed method is evaluated on a 400 kV 50 Hz power system. Both sides of this network are two sources that represent two areas which are connected together. The source data at both sides of the transmission line is shown in Table I. Frequency-dependent (phase) model is used for detailed simulation of transmission line. Transmission line data is achieved from a real 400 kV single-circuit transmission line with a length of 250 km. this data is presented in Table II.

Following this line of 250 kilometers, a line with a length of 100 km and then a line with length of 50 km with the same profile was considered. To evaluate the impact of TCSC location on the issue of protection, this element was placed in two locations, first in the middle of the line 1 (125 km of line 1) and second, in Seventy-five percent of the line 1 (187.5 km of line 1), which in continuation of this paper for the convenience of readers will be referred to as scenario 1 and scenario 2, respectively. Three protection zones have been defined for distance relay, first till 80 percent of the first line, second till 50 percent of the second line and third till 25 percent of the third line.

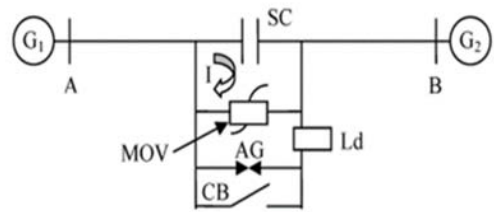

Fig 1. TCSC simulation model [13]

TABLE I. SOURCE DATA AT BOTH ENDS

| Parameters | Value |
|---|---|
| Source Voltage | 400 kV |
| Frequency | 50 Hz |
| Positive sequence impedance | 1.31+j15 Ω |
| *Zero sequence impedance* | *2.33+j26.6 Ω* |

TABLE II. TRANSMISSION LINE DATA

| Data type | value | Data type | value |
|---|---|---|---|
| Arrangement of wires | Horizontal | Vertical distance of conductors | 15.45 m |
| Line length | 250 km | SAG for all conductors | 14 m |
| Number of bundle | 2 | Number of ground wires | 2 |
| Bundle spacing | 45 cm | Height of ground wires above the lowest conductor | 9.36 m |
| Conductor DC resistance | 0.0553 ohm/km | Ground wire DC resistance | 1.463 ohm/km |
| Conductor GMR | 1.2161 cm | Ground wire radius | 0.2445 cm |
| Height of conductors | 41.46 m | Spacing between ground wires | 18.70 m |

In scenario one which TCSC is placed in the middle of the first line, four fault locations are assumed: in 50 km (first zone, before TCSC), in 150 km (first zone, after TCSC), in 250 km (second zone), in 325 km (third zone). In scenario 2, three fault locations are assumed: in 100 km (first zone), in 250 km (second zone), in 325 km (third zone).

In order to evaluate the feasibility of proposed method under various system condition, different kinds of fault type, fault resistance, fault inception angle (FIA), load angle (δ), generator impedance, compensation percentage and fault location, which include 28800 cases for scenario 1 and 21600 cases for scenario 2, have been analyzed utilizing multiple run in PSCAD. Table III shows considered cases. 25%, 50%, and 75% are chosen TCSC compensation percentages for 250 km transmission line, their equivalent value for SC capacitor is 120.45, 60.225 and 40.15 µf respectively.

Randomly chosen cases for the training process of SVM are 3600 cases for scenario 1 and 2700 cases for scenario 2. These cases are only 12.5% of the total data set and are shown in Table IV. The Gaussian RBF kernel has been used for training and testing process in SVM.

*B. Classification with One-Against-All method (OAA)*

.There are various ways to use SVM to classify data into more than two classes. In OAA assuming k classes, k SVM is trained for each class with the samples of that class as positive samples and all other samples as negatives.

*1) Analysis of scenario 1*

Getting a proper output from SVM can be achieved through metaheuristic algorithms or trial and error learning [13]. In order to utilize trial and error learning different values for regularization parameter (C) and gamma (g) (RBF kernel parameter) should be evaluated on the classification of training data in order to choose the best ones, then the result is used to classify the test data. For the range of these parameters, results in reference [13] was a great help. The best result is achieved in 21096 and 14.5 for C and g, respectively. With these values, 83.7% of success is achieved in fault zone detection.

*2) Analysis of scenario 2*

In result analysis of scenario two, it could be seen that the most number of correct fault zone detection was reached with C=1080000 and g=38.4. The success rate was 69.7%, which was less than scenario one. In section IV causes of this lower success rate will be discussed.

TABLE III. CONSIDERED CASES FOR SYSTEM ANALYSIS

| Case number | Number of fault cases for scenario 1 | Number of fault cases for scenario 2 | Parameters ||||||| |
|---|---|---|---|---|---|---|---|---|---|---|
| | | | %$Z_{G1}$ | %$Z_{G2}$ | %$Xc$ | $R_f$ | FIA | δ | Fault types | Fault locations |
| 1 | 5760 | 4320 | 100 | 100 | 25,50 & 75 | 0,5 & 25,50 | 0,45 & 81,117 | 10,20 & 30 | 10 | 4 and 3 location for scenario 1 and scenario 2 respectively |
| 2 | 5760 | 4320 | 100 | 75 | | | | | | |
| 3 | 5760 | 4320 | 100 | 125 | | | | | | |
| 4 | 5760 | 4320 | 75 | 100 | | | | | | |
| 5 | 5760 | 4320 | 125 | 100 | | | | | | |
| total number of cases in scenario 1= 5*5760=28800 |||||||||||
| total number of cases in scenario 2= 5*4320=21600 |||||||||||

TABLE IV. CHOSEN CASES FOR TRAINING PROCESS OF SVM

| Case number | Number of fault cases for scenario 1 | Number of fault cases for scenario 2 | Parameters ||||||| 
|---|---|---|---|---|---|---|---|---|---|---|
| | | | %$Z_{G1}$ | %$Z_{G2}$ | %$X_c$ | $R_f$ | FIA | δ | Fault types | Fault locations |
| 1 | 720 | 540 | 100 | 100 | 50 | 0,5 & 50 | 0,45 & 117 | 10 & 30 | 10 | 4 and 3 location for scenario 1 and scenario 2 respectively |
| 2 | 720 | 540 | 100 | 75 | 50 | | | | | |
| 3 | 720 | 540 | 100 | 125 | 50 | | | | | |
| 4 | 720 | 540 | 75 | 100 | 50 | | | | | |
| 5 | 720 | 540 | 125 | 100 | 50 | | | | | |
| total number of cases in scenario 1= 5*720=3600 ||||||||||| 
| total number of cases in scenario 2= 5*540=2700 ||||||||||| 

*C. Classification with One-Against-One method (OAO)*

For this method, a table should be established, and each SVM should cast its vote on a class between two classes. The class that achieves the highest number of votes is chosen by the combined classifiers. Possible scenarios are shown in Table V.

*1) Analysis of scenario 1*

The maximum number of correct detection of fault zone using OAO in scenario 1 was 20629 cases which happened in C=10000 and g=13.1. This number is equivalent to 81.8% of success.

Following up on OAO method, with analysis of results, one can decide for cases that the table cannot vote for their class. After checking the results, Table V was replaced with Table VI.

The maximum number of correct detections of fault zone using Table VI was 21239, which happened in C=10 and g=10.4. As expected, the success rate increased to 84.2%.

*2) Analysis of scenario 2*

The maximum number of correct detection of fault zone using OAO in scenario 2 was 11802 cases which happened in C=10 and g=20. This number is equivalent to %62.2 of success. After replacing Table V by Table VI number of correct cases increased to 123399 in C=10 and g=21.3, which is equivalent to 65.6% of success.

*D. Classification with OAA method in the modified training set*

As can be seen, a substantial percentage of success was not achieved for scenario 1 and 2 with both methods. One of the effective parameters to achieve a good result in classification is having a proper and right amount of training data set. One cannot expect an algorithm to predict properly in unseen areas. It is usual to put 70% of the data set into the training set and 30% to the test set. Until now, only 12.5% of the data set was used in the training set. Cases, which their fault zone was identified incorrectly, were analyzed. A high proportion of wrong predictions happened in data with load angles (δ) equal to 20, which was not included in the training data set. Therefore, random data sets with load angles (δ) equal to 20 were added to the training set. Table VII and Table VIII Shows the added data for scenario 1 and scenario 2, respectively. Consider that now 15.3% of total data is used as the training data set.

*1) Analysis of scenario 1*

The maximum number of correct detection of fault zone using OAA in the modified training set was 22993 cases which happened in C=10 and g=8.3. Modified training set increased the success rate from 83.7% to 94.2%.

*2) Analysis of scenario 2*

The maximum number of correct detection of fault zone using OAA in the modified training set was 15880 cases which happened in C=10 and g=14.5. Modified training set increased the success rate from 69 .7% to 86.2%.

TABLE V. POSSIBLE SCENARIOS IN VOTING

| Groups | Possible conditions ||||||||
|---|---|---|---|---|---|---|---|---|
| 1,2 | 1 | 1 | 1 | 1 | 0 | 0 | 0 | 0 |
| 1,3 | 1 | 1 | 0 | 0 | 1 | 1 | 0 | 0 |
| 2,3 | 0 | 1 | 0 | 1 | 0 | 1 | 1 | 0 |
| result | 1 | 1 | 3 | 0 | 0 | 2 | 2 | 3 |

TABLE VI. MODIFIED SCENARIOS IN VOTING

| Groups | Possible conditions ||||||||
|---|---|---|---|---|---|---|---|---|
| 1,2 | 1 | 1 | 1 | 1 | 0 | 0 | 0 | 0 |
| 1,3 | 1 | 1 | 0 | 0 | 1 | 1 | 0 | 0 |
| 2,3 | 0 | 1 | 0 | 1 | 0 | 1 | 1 | 0 |
| result | 1 | 1 | 3 | 3 | 2 | 2 | 2 | 3 |

TABLE VII. ADDED DATA TO TRAINING SET IN SCENARIO 1

| Case number | Number of fault cases | Parameters | | | | | |
|---|---|---|---|---|---|---|---|
| | | %$Z_{G1}$ | %$Z_{G2}$ | %$X_C$ | $R_f$ | FIA | $\delta$ |
| 1 | 160 | 100 | 100 | 25 & 50 | 50 | 81 & 117 | 20 |
| 2 | 160 | 100 | 75 | | | | |
| 3 | 160 | 100 | 125 | | | | |
| 4 | 160 | 75 | 100 | | | | |
| 5 | 160 | 125 | 100 | | | | |
| Total number=5*160=800 | | | | | | | |

TABLE VIII. ADDED DATA TO TRAINING SET IN SCENARIO 2

| Case number | Number of fault cases | Parameters | | | | | |
|---|---|---|---|---|---|---|---|
| | | %$Z_{G1}$ | %$Z_{G2}$ | %$X_C$ | $R_f$ | FIA | $\delta$ |
| 1 | 120 | 100 | 100 | 25 & 50 | 0 & 50 | 81 | 20 |
| 2 | 120 | 100 | 75 | | | | |
| 3 | 120 | 100 | 125 | | | | |
| 4 | 120 | 75 | 100 | | | | |
| 5 | 120 | 125 | 100 | | | | |
| Total number=5*120=600 | | | | | | | |

TABLE IX. MODIFIED SCENARIOS IN VOTING FOR SECTION E

| Groups | Possible conditions | | | | | | | |
|---|---|---|---|---|---|---|---|---|
| 1,2 | 1 | 1 | 1 | 1 | 0 | 0 | 0 | 0 |
| 1,3 | 1 | 1 | 0 | 0 | 1 | 1 | 0 | 0 |
| 2,3 | 0 | 1 | 0 | 1 | 0 | 1 | 1 | 0 |
| result | 1 | 1 | 3 | 2 | 3 | 2 | 2 | 3 |

*E. Classification with OAO method in the modified training set*

  *1) Analysis of scenario 1*

In this case, the best result is achieved in C=10 and g=12.1. With these values, the success rate is increased to 86.7%. After analyzing the results, Table V was replaced with Table IX as the reference voting table. The results showed 91.6% success rate in C=10 and g=12.1.

  *1) Analysis of scenario 2*

The results showed 76.1% success rate in C=100000 and g=15.3, which is equivalent to 13926 correct fault zone detections. After analyzing the results, Table V was replaced with Table IX as the reference voting table. The results showed 78.2% success rate in C=100000 and g=15.4, which is equivalent to 14318 correct fault zone detections.

## IV. DISCUSSION ON THE RESULTS

According to the results, the highest success rate is achieved in scenario 1 (installing TCSC in the middle of the first line). This success rate (94.2%) was reached with OAA method on the modified training set. In scenario two, the highest success rate was 86.7%, which was also reached with OAA method on the modified training set. Thus, the prominent role of selecting a proper training data set had become evident. To investigate why scenario 2 had less success rate than scenario 1, cases with the wrong detection of the fault zone in both scenarios were analyzed in Table X and Table XI. By comparing these two tables, it can be concluded that an increase in wrong zone detections happened in zone 2 and 3, whereas in zone 1 it almost remained constant. In other words, SVM made more mistakes because pushing the TCSC forward increased its area of influence.

## V. CONCLUSIOIN

In this paper, a new method for fault zone detection of distance relay in FACTS-based transmission lines has been proposed. For investigating the effects of TCSC location on fault zone detection of distance relay, two places, one in fifty percent of line length and the other in seventy-five percent of line length, have been considered as two scenarios. In total, four cases have been considered for each scenario. For scenario 1, with OAA method, 83.7% of success, and with OAO method, 84.4% of success have been achieved. After analyzing the result and increasing the training set, with OAA method, 94.2% of success and with OAO method, 91.6% of success have been achieved. For scenario 2, with OAA method, 69.7% of success, and with OAO method, 65.6% of success have been achieved. After analyzing the result and increasing the training set, with OAA method, 86.7% of success and with OAO method, 78.2% of success have been achieved. Results showed that in solving this type of problem, ultimately, the OAA method has a comparative advantage over the OAO method. With the OAA method, 94.2% and 86.7% of success have been achieved in scenario 1 and 2, respectively. It was shown that installing TCSC in the middle of the first line leads to better results by decreasing its influence on zone 2 and 3 of the distance relay. Transmission lines are vital parts of power system. Improving their protection is necessary for reliability of the network. This paper presents a novel approach to increase the reliability of the system by improving the protection of transmission lines in fault instance and minimizing the damage to other network equipment. Rapid detection and isolation of a faulty transmission line, protect the network from its possible harmful effects

TABLE X. NUMBER OF WRONG PREDICTIONS IN SCENARIO 1

| Predicted zone / Real zone | 1 | 2 | 3 | Total number of wrong prediction in each zone |
|---|---|---|---|---|
| 1 |  | 99 | 211 | 310 |
| 2 | 249 |  | 512 | 761 |
| 3 | 193 | 143 |  | 336 |

TABLE XI. NUMBER OF WRONG PREDICTIONS IN SCENARIO 2

| Predicted zone / Real zone | 1 | 2 | 3 | Total number of wrong prediction in each zone |
|---|---|---|---|---|
| 1 |  | 91 | 237 | 328 |
| 2 | 829 |  | 613 | 1442 |
| 3 | 381 | 283 |  | 644 |